\documentclass{elsart}

\usepackage{graphicx}

\usepackage{amssymb}

\begin{document}

\begin{frontmatter}



\title{Improvement of the robustness on geographical networks by
adding shortcuts}


\author[JAIST]{Yukio Hayashi}, 
\author[KNCT]{Jun Matsukubo}

\address[JAIST]{Japan Advanced Institute of Science and Technology, 
Ishikawa 923-1292, Japan}
\address[KNCT]{Kitakyusyu National College of Technology, 
Fukuoka 802-0985, Japan}

\begin{abstract}
In a topological structure affected by geographical constraints on
liking, the connectivity is weakened by constructing local stubs with
small cycles, a something of randomness to bridge them is crucial for
the robust network design.
In this paper, we numerically investigate the effects of adding
shortcuts on the robustness in geographical scale-free network models
under a similar degree distribution to the original one.
We show that a small fraction of shortcuts is highly contribute to
improve the tolerance of connectivity especially for the intentional
attacks on hubs.  The improvement is equivalent to the effect by 
fully rewirings without geographical constraints on linking.
Even in the realistic Internet topologies,
these effects are virtually examined.
\end{abstract}

\end{frontmatter}

\section{Introduction}
In many social, technological, and biological networks, 
there exist several common topological characteristics 
which can be emerged 
by simple generation mechanisms \cite{Barabasi02,Buchanan02},  
and a small fraction of networks crucially influences the communication
properties for the routing and the robustness of connectivity.
One of the sensational facts is the {\bf small-world} (SW) phenomenon:
each pair of nodes are connected through relatively small number of hops 
to a huge network size defined by the total number of nodes.
This favorable phenomenon has been explained by the SW model with 
a small fraction of random rewirings
on a one-dimensional lattice \cite{Watts98}.
Another fact is the {\bf scale-free} (SF) structure that follows a power-law
degree distribution $P(k) \sim k^{- \gamma}$, $2 < \gamma < 3$,
consists of many nodes with low degrees and a few hubs with high degrees.
The heterogeneous networks are 
drastically broken into many isolated
clusters, when only a small fraction of high degree nodes are removed 
as the intentional attacks \cite{Albert00,Cohen00a}.
However, the SF structure is robust against random failures
\cite{Albert00,Cohen00b}, 
and well balanced in the meaning of both economical and
efficient communication by small number of hops in a connected network
as few links as possible \cite{Cancho03}.

On the other hands, real complex networks, 
such as power-grid, airline flight-connection, and the Internet, 
are embedded in a metric space,
and long-range links are restricted \cite{Yook02,Gastner06}
for economical reasons.
Based on the connection probability according to a penalty of distance 
$r$ between two nodes 
and on random triangulation, 
the generation mechanisms of geographical SF networks have been
proposed in {\bf lattice-embedded scale-free} (LESF) \cite{Avraham03}
and {\bf random Apollonian} (RA) \cite{Zhou05,Doye05} network models.
Unfortunately, the vulnerability of connectivity
has been numerically found in both networks
\cite{Huang05a,Huang05b,Hayashi06b}.
Moreover, it has been theoretically predicted \cite{Huang05a}
in a generating function approach to more
general networks with any degree distribution
that a geographical constraint on linking decreases the
tolerance to random failures, 
since the percolation threshold is increased by 
the majority of small-order cycles that locally 
connected with a few hops.
As the smallest-order,
triangular cycles tend to be particularly constructed by a 
geographical constraint.

In contrast, it has been suggested that 
higher-order cycles connected with many hops
improve the robustness in the theoretical analysis 
on a one-dimensional SW model modified by adding 
shortcuts between two nodes out of the connected neighbors 
\cite{Newman00a,Newman00b}.
Similarly, the expected delivery time of any decentralized search
algorithm without global information is decreased 
on a two-dimensional lattice whose each node has a shortcut with the 
connection probability proportional to the power of distance 
$r^{- \alpha}$, $\alpha > 0$ \cite{Kleinberg00}.
These results support the usefulness of shortcuts for maintaining 
both the robust connectivity and the communication efficiency, 
however the network structures are almost regular and far from the
realistic SF.
Recently, 
it has been numerically shown \cite{Hayashi06b} 
that the robustness is improved 
by fully random rewirings under a same degree distribution \cite{Maslov04} 
in typical geographical network models: 
{\bf Delaunay triangulation} (DT) \cite{Imai00,Okabe00}, 
RA, and {\bf Delaunay-like scale-free} 
(DLSF) networks \cite{Hayashi06b}.
Instead of rewirings, 
we expect the shortcut effect on the improvement of 
robustness in such geographical SF networks.
Adding shortcuts is practically more natural rather than rewirings, 
because the already constructed links are not wastefully discarded.
Thus, we investigate how large connected component still remains
at a rate of removed nodes as the random failures and the targeted
attacks on hubs in the geographical SF networks with shortcuts,
comparing the original ones without shortcuts.
We show that a small quantity of geographical randomness highly
contributes to maintain both the communication efficiency and the
robustness under almost invariant degree distributions
to the original ones.
It is not trivial that the improvement of robustness is equivalent to
the effect by fully random rewirings.

The organization of this paper is as follows.
In Sec. \ref{sec2}, we briefly 
introduce the geographical 
networks based on planar triangulation and embedding in a lattice under a
given power-law degree distribution.
In Sec. \ref{sec3}, we numerically investigate 
the effects of shortcuts on the optimal paths in two measures of
distance/hop and the robustness in the geographical networks.
In particular, we show that 
a degree cutoff enhances the improvement of the error and attack
tolerance.
Moreover, we virtually examine the effects for realistic data of the
Internet topologies.
Finally, in Sec. \ref{sec4} we summarize these results.

\section{Geographical SF networks} \label{sec2}
\subsection{Planar network models}
Planar networks without crossing of links 
are suitable for efficient geographical routings, 
since we can easily find the shortest path from a set of edges of the
faces that intersect the straight line between the source and terminal.
In computer science, online routing algorithms \cite{Bose04}
that guarantee delivery of messages using only
local information about positions of the source, terminal, and the 
adjacent nodes to a current node are well-known. 
As a connection to SF networks, 
we consider Delaunay triangulation (DT) and random Apollonian (RA)
network models based on planar triangulation of a polygonal region.
DT is the optimal planar triangulation in some geometric criteria
\cite{Imai00} with respect to the maximin angle and 
the minimax circumcircle of triangles on a two-dimensional space.
In addition, 
the ratio of the shortest path length is bounded by a constant factor
to the direct Euclidean distance between any source and terminal
\cite{Keil92}, 
while RA network belongs to both SF and planar networks
\cite{Doye05,Zhou05},
however long-range links inevitably appear near the edges of
an initial polygon.
To reduce the long-range links, 
Delaunay-like scale-free (DLSF) network 
has been proposed \cite{Hayashi06b}.

On the preliminaries, just like overhead highways,
we add shortcuts between randomly chosen two nodes 
excluding self-loops and multi-links after constructing
the above networks.
For adding shortcuts, the routing algorithm can be extended 
as mentioned in Appendix 1.
Note that the added shortcuts contribute to create
some higher-order cycles which consists of a long path and the overhead
bridge in the majority of triangular cycles.
The original degree distributions without shortcuts 
follow a power-law with the exponent nearly 3 in RA, 
log-normal in DT, and power-law with an exponential
cutoff in DLSF networks \cite{Hayashi06b}.
Note that the lognormal distribution has an unimodal shape as similar to 
one in Erd\"{o}s-Renyi random networks.
Thus, RA and DLSF networks are vulnerable because of double 
constraints of planarity and geographical distances on the linkings, 
but DT networks are not so.
We have confirmed that the degree distributions 
have only small deviation from the original ones 
at shortcut rate up to the amount of 
$30 \%$ of the total links.

\subsection{Lattice-embedded SF networks}
Let us consider a $d$-dimensional lattice of size $R$ with the periodic  
boundary conditions.   
The LESF network model \cite{Avraham03} combines 
the configuration model \cite{Newman01} for any degree distribution 
with a geographical constraint on a lattice.
Although the homogeneous positioning of nodes differs from a realistic
spacial distribution such as in the Internet routers according to the
population density \cite{Yook02}, it has been studied as a fundamental 
spacial model.
Note that the spatial distribution of nodes is restricted on the
regular lattice, some links are crossed,
therefore LESF networks are not planar.

In the following simulation, 
we assign a degree taken from the distribution $P(k) \sim k^{- 3}$
to each node on a two-dimensional lattice of the network size 
$N = 32 \times 32$, where $\gamma = 3$, $d = 2$, and $R = 32$.
The networks have the average numbers 
$M \approx 1831$ of the total links 
at $A = 1$ and $M \approx 2673$ at $A = 3$
for comparison.
The case of $A \rightarrow \infty$ is equivalent to the 
Warren-Sander-Sokolov model \cite{Warren02} whose degree distribution
follows a pure power-law, 
however a cutoff is rather natural in real networks with something of
constraints on linkings \cite{Amaral00}.
As similar to the previously mentioned planar networks, 
there are little deviation form the original power-law 
distributions with strong and weak cutoffs
at $A=1$ and $A=3$, respectively.
The detailed configuration procedures for RA, DT, DLSF, and LESF
networks are summarized in Appendix 2.

\section{Shortcut effects} \label{sec3}
\subsection{Shortest distance and minimum hops}
For the shortcut rates from $0\%$ to $30\%$, 
we investigate four combinations of distance/hops and two kinds of 
the optimal paths 
with respect to the shortest {\bf distance} 
and the minimum number (or called {\bf length}) of hops: 
the average distance 
$\langle D \rangle$ on the shortest paths, 
the distance $\langle D' \rangle$ on the paths of the minimum hops, 
the average number of hops $\langle L \rangle$ on these paths, 
and the number of hops 
$\langle L' \rangle$ on the shortest paths between any two nodes
in the geographical networks.
The prime denotes the cross relation to the case of no prime in 
the combinations of the measures and the two kinds of paths.
The distance is defined by a sum of link lengths on the path, 
and the average means a statistical ensemble over the optimal paths 
in the above two criteria.
Note that the shortest path and the path of the minimum hops may be
distinct, these are related to 
the link cost or delay and the load for transfer of a message.
It is better to shorten both the distance and the number of hops, 
however the constraints are generally conflicted.

\begin{figure}[htb]
  \begin{minipage}[htb]{.47\textwidth}
    \includegraphics[height=48mm]{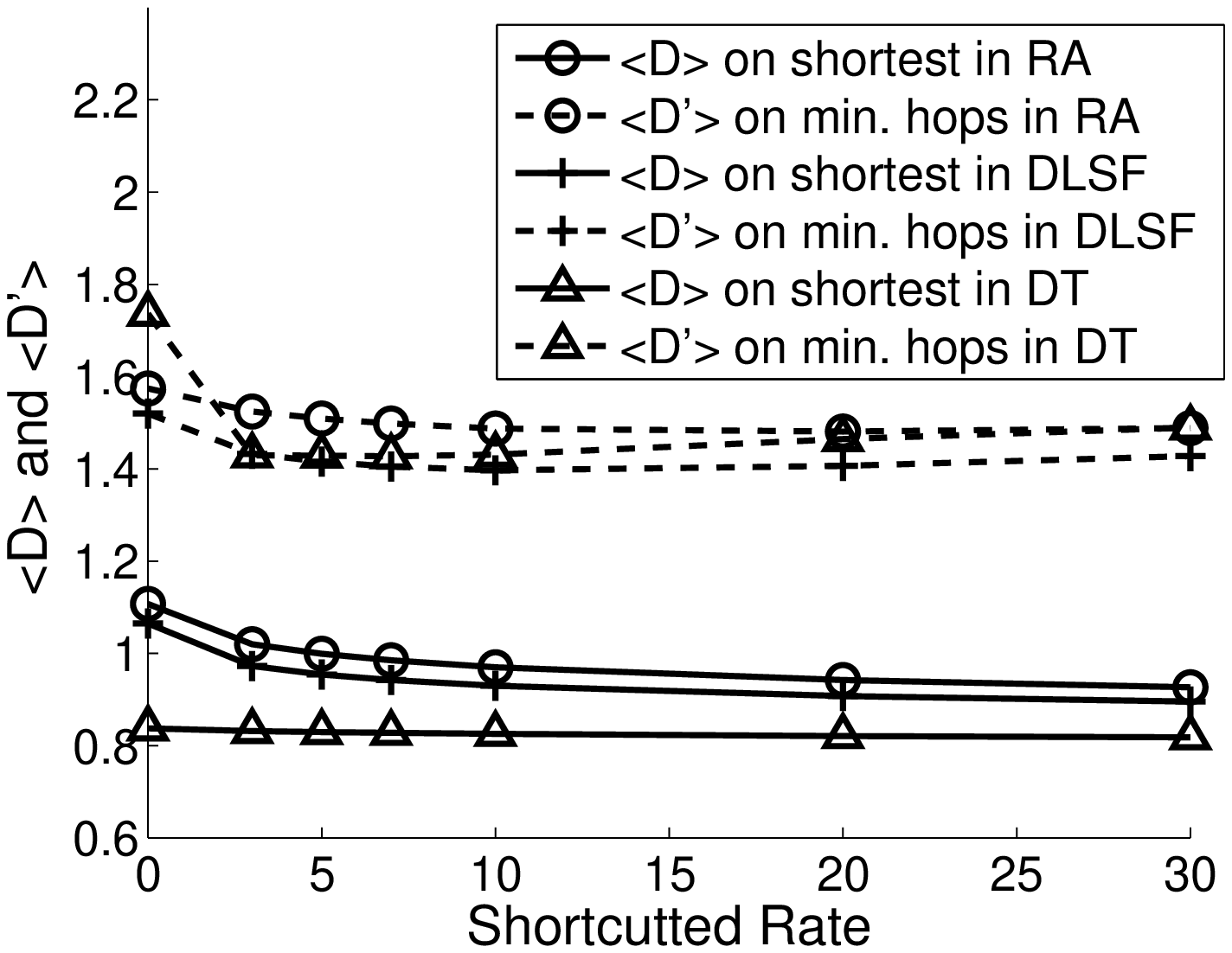}
    \begin{center} (a) average distance \end{center}
  \end{minipage}
  \hfill
  \begin{minipage}[htb]{.47\textwidth}
    \includegraphics[height=48mm]{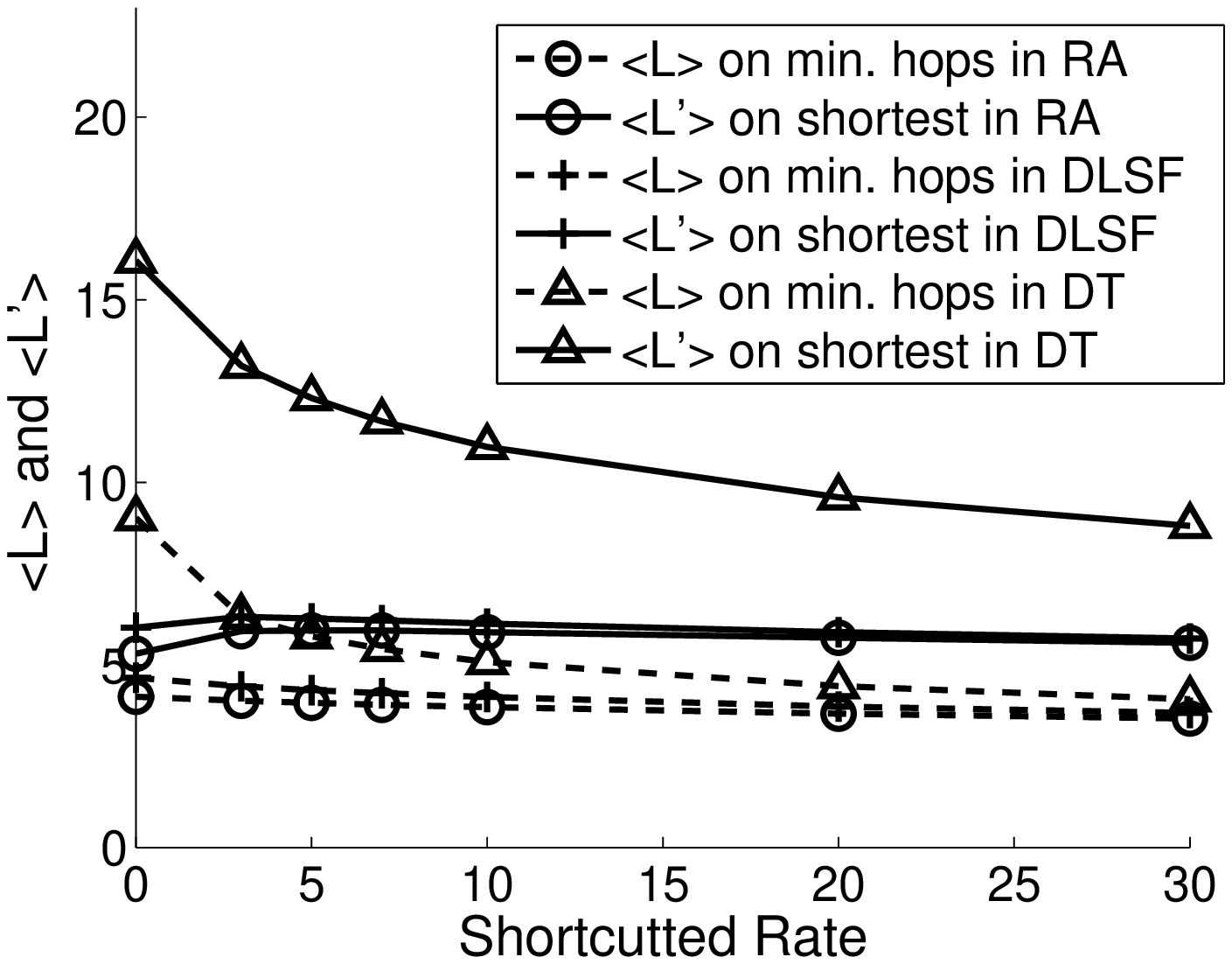}
    \begin{center} (b) average number of hops \end{center}
  \end{minipage}
  \caption{The average distance and the number of hops
 on two kinds of the optimal paths in DT (triangle), RA (circle), 
 DLSF (plus) networks.
 Solid lines guide the decreasing or increasing of
 $\langle D \rangle$ and $\langle L' \rangle$ on the shortest paths,
 dashed lines guide that of
 $\langle D' \rangle$ and $\langle L \rangle$
 on the paths of the minimum hops.}
  \label{fig_D_L}
\end{figure}

\begin{figure}[htb]
  \begin{minipage}[htb]{.47\textwidth}
    \includegraphics[height=48mm]{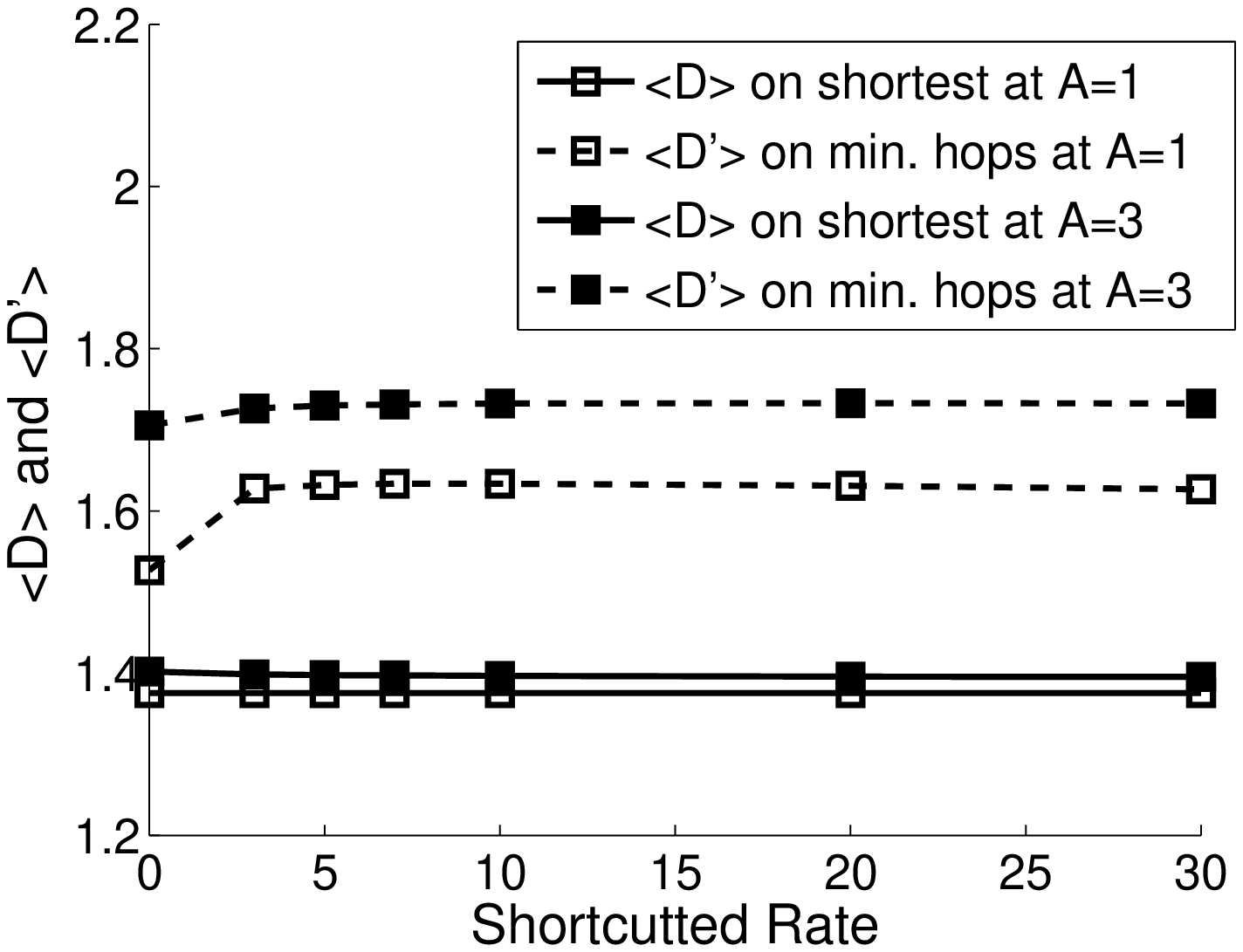}
    \begin{center} (a) average distance \end{center}
  \end{minipage}
  \hfill
  \begin{minipage}[htb]{.47\textwidth}
    \includegraphics[height=48mm]{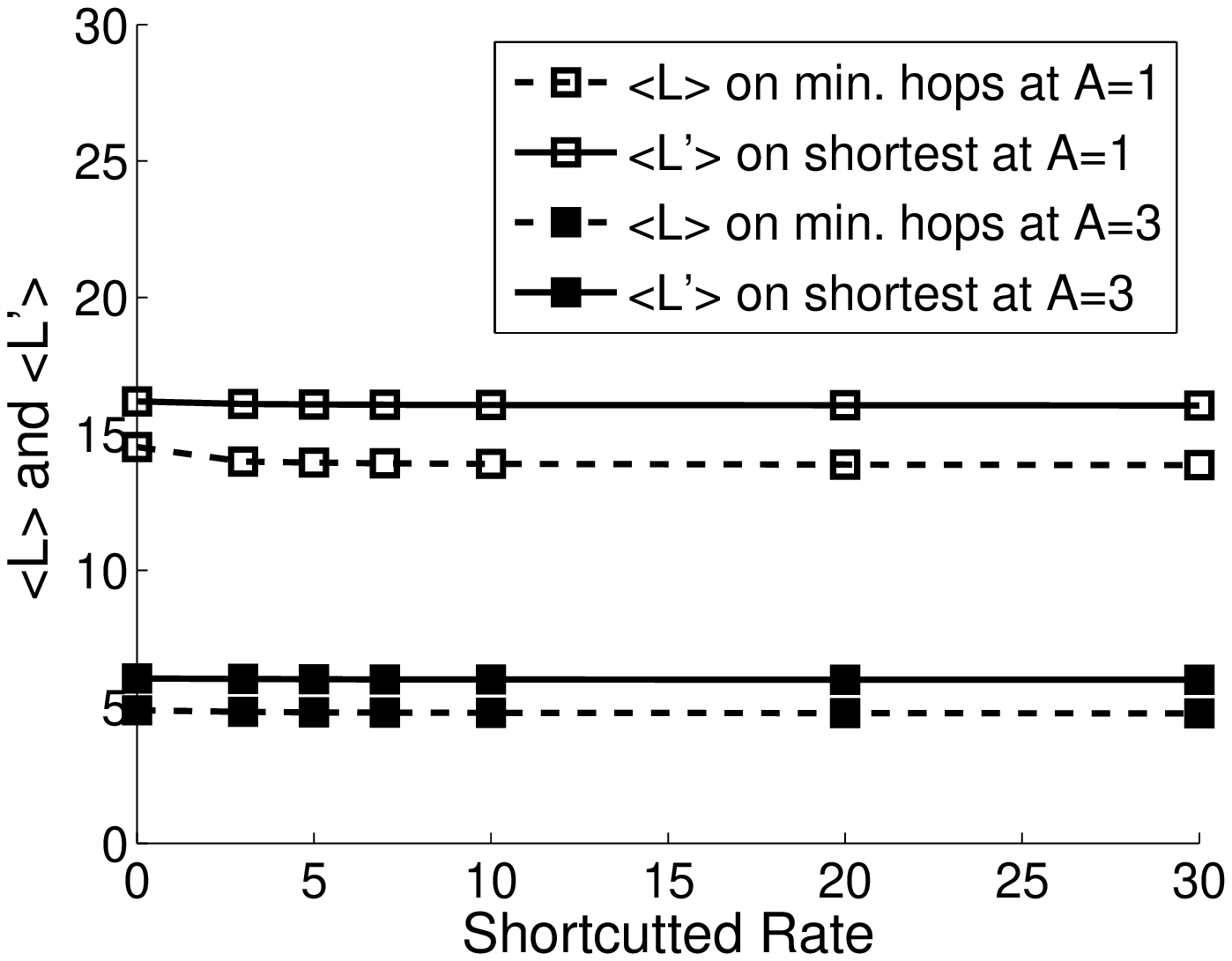}
    \begin{center} (b) average number of hops \end{center}
  \end{minipage}
  \caption{The average distance and the number of hops
 on two kinds of the optimal paths in LESF networks 
 at $A = 1$ (open rectangle) and $A = 3$ (filled rectangle).
 Solid lines guide the decreasing or increasing of
 $\langle D \rangle$ and $\langle L' \rangle$ on the shortest paths,
 dashed lines guide that of
 $\langle D' \rangle$ and $\langle L \rangle$
 on the paths of the minimum hops.}
  \label{fig_SFL_D_L}
\end{figure}

In the original networks without shortcuts, 
we note the tendencies \cite{Hayashi06b}: 
RA networks have a path connected by a few hops 
but the path length tend to be long including 
some long-range links,
while DT networks have a zig-zag path connected by many hops 
but each link is short,  in addition, 
DLSF networks have the intermediately balanced properties.
Figure \ref{fig_D_L} shows numerical results in adding shortcuts
to the planar networks.
We find that,
from the solid and dashed lines in Figs. \ref{fig_D_L}(a)(b),
the average distance $\langle D \rangle$ and 
the number of hops $\langle L \rangle$
become shorter as increasing the shortcut rate.
In particular, the shortcuts are effective for the distance in RA and
DLSF (solid lines marked with circles and pluses) networks, 
and also for the number of hops in DTs (dashed line marked with triangles).
On another measures of $\langle D' \rangle$ and $\langle L' \rangle$,
the dashed lines for DTs (marked with triangles) and RA (with circles) 
networks in Fig. \ref{fig_D_L}(a) 
and the solid lines for RA (with circles)
and DLSF (with pluses) networks in Fig. \ref{fig_D_L}(b) approach to
each other.
Thus, the shortcuts even around $10 \%$ decrease 
both  $\langle D \rangle$ and $\langle L \rangle$,
and maintain small $\langle D' \rangle$ and $\langle L' \rangle$.
On the other hand, 
as shown in Fig. \ref{fig_SFL_D_L},
the average distance and the number of hops 
are almost constant in LESF networks.
Probably, 
the links emanated from hubs already 
act as shortcuts on the lattice.
These results are obtained from 
ensembles over 100 realizations for each network model.

\subsection{Robustness of connectivity}
\begin{figure}[htb]
  \begin{minipage}[htb]{.3\textwidth}
    \includegraphics[height=37mm]{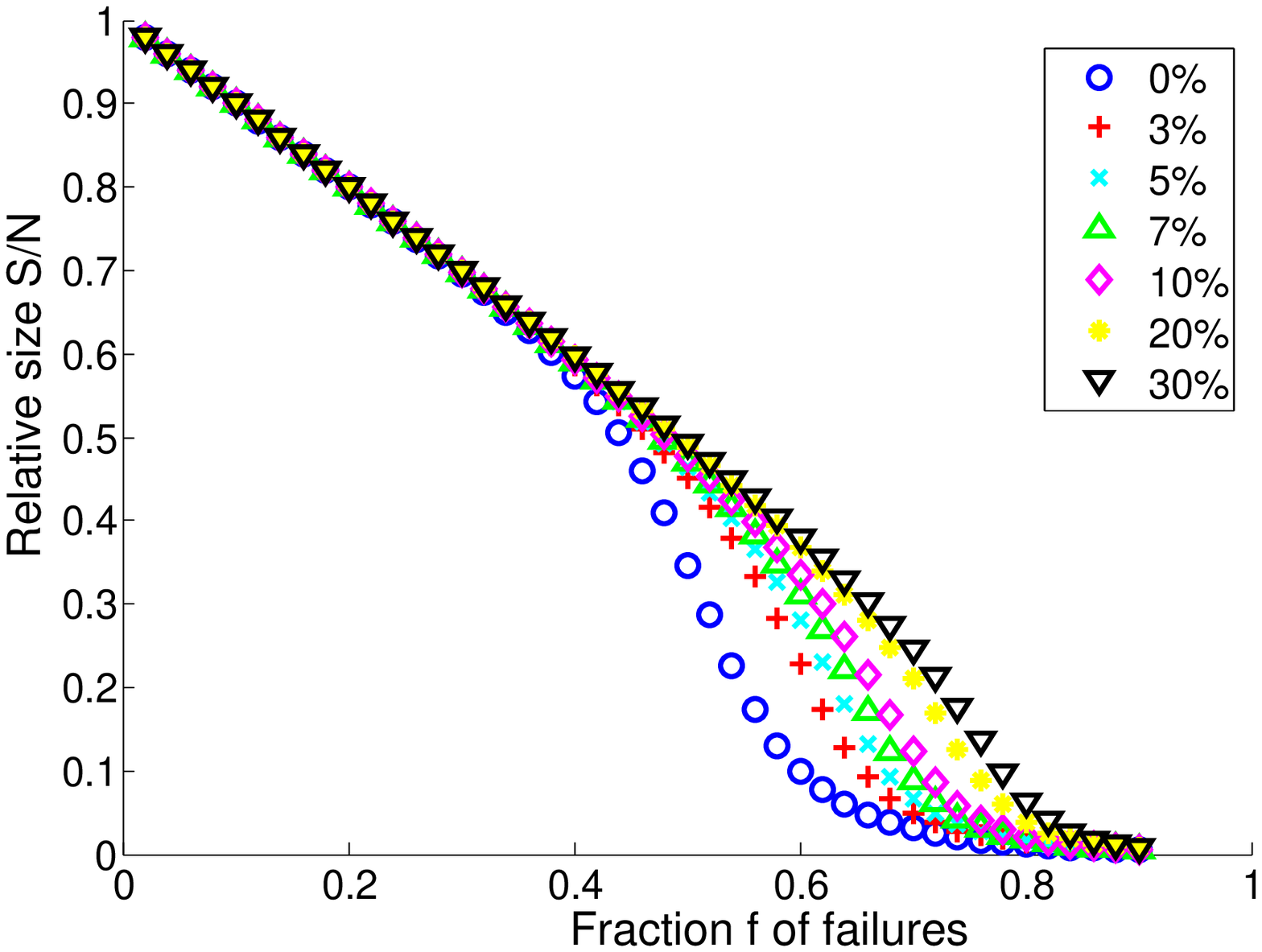}
    \begin{center} (a) DT \end{center}
  \end{minipage}
  \hfill
  \begin{minipage}[htb]{.3\textwidth}
    \includegraphics[height=37mm]{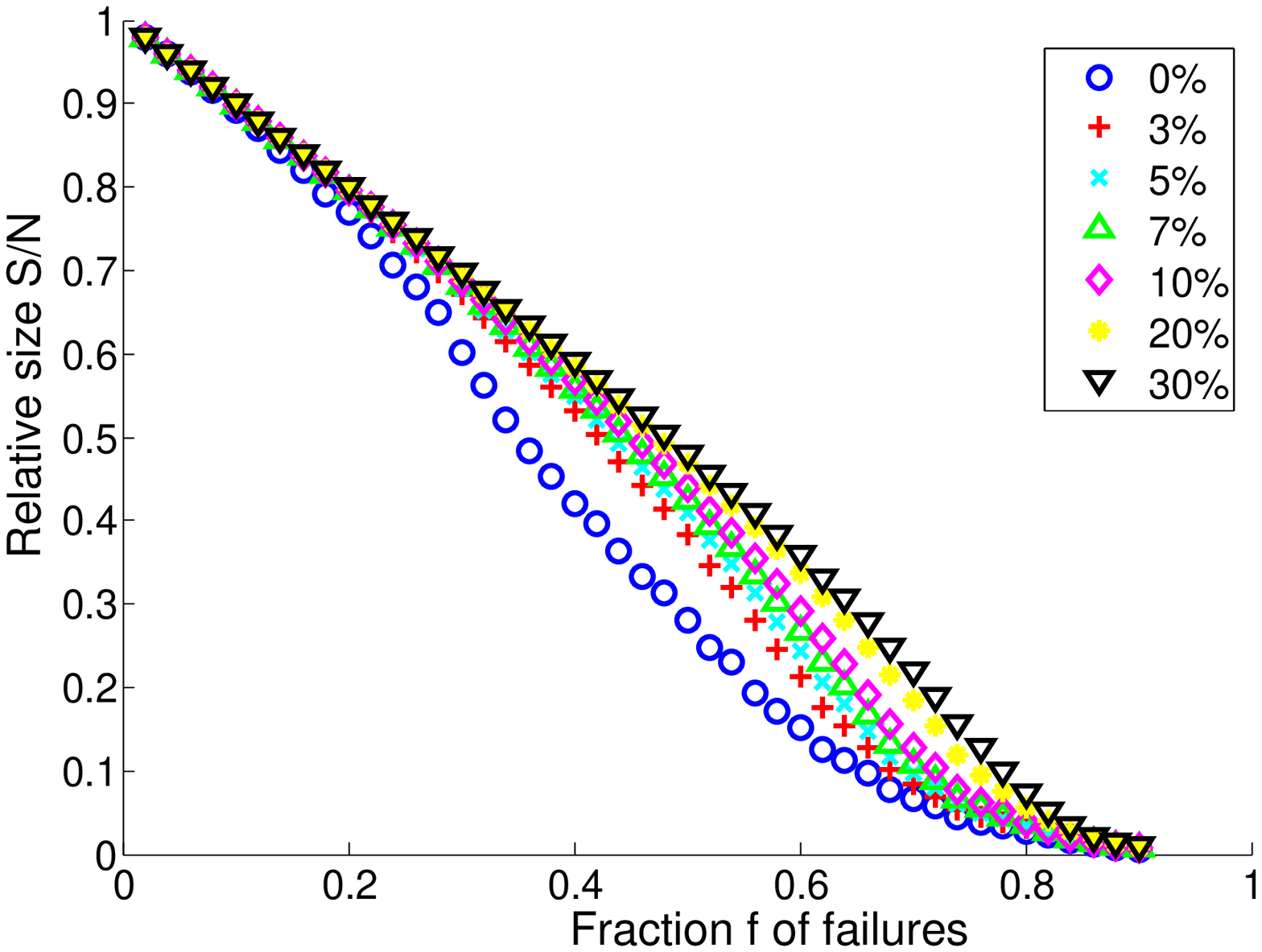}
    \begin{center} (b) RA \end{center}
  \end{minipage}
  \hfill
  \begin{minipage}[htb]{.3\textwidth}
    \includegraphics[height=37mm]{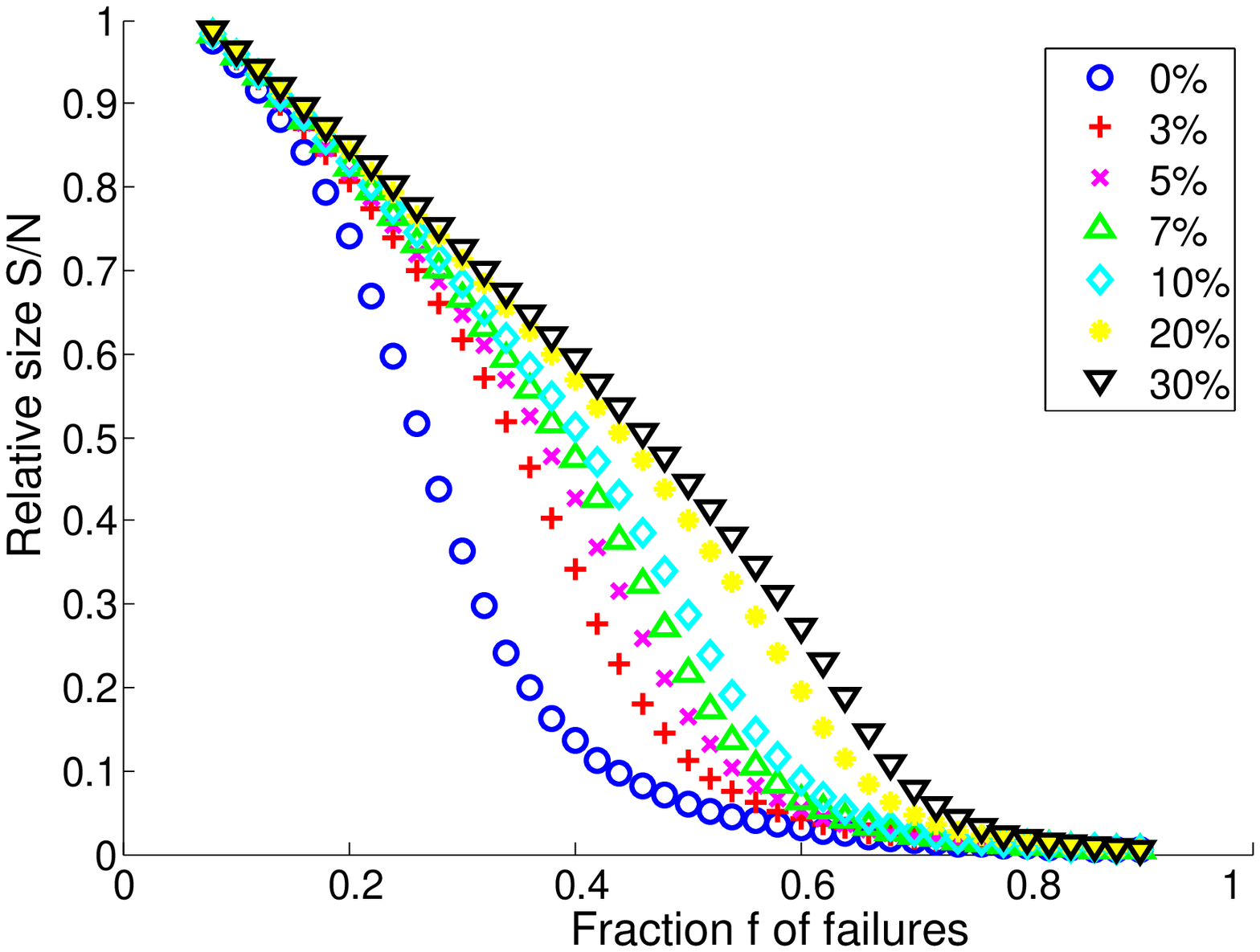}
    \begin{center} (b) LESF at $A=1$
   \end{center}
  \end{minipage}
  \caption{(Color online) Typical results of 
 the relative size $S/N$ of the GC against random
 failures in (a) DT, (b) RA, and (c) LESF ($A=1$) networks 
 at the shortcuts rates in legend.}
  \label{fig_GC_rand}
\end{figure}

The fault tolerance and attack vulnerability are known as the typical
properties of SF networks \cite{Albert00},
which are further affected by geographical constraints.
We investigate the tolerance of connectivity in the giant component (GC)
of the geographical networks with shortcuts 
comparing with that of the original ones without shortcuts.
The size $S$ of GC and the average size $\langle s \rangle$ of isolated
clusters are obtained from 
ensembles over 100 realizations for each network model.
Figure \ref{fig_GC_rand} shows the typical results 
that a small fraction of shortcuts suppresses the breaking
of the GC against random failures.
It seems to be enough in less than $10 \%$.
In other DLSF, LESF ($A=3$) networks, 
the effects are similar.
Thus, the added shortcuts strengthen the tolerance in comparison with 
each original network.

\begin{figure}[htb]
  \begin{minipage}[htb]{.3\textwidth}
    \includegraphics[height=37mm]{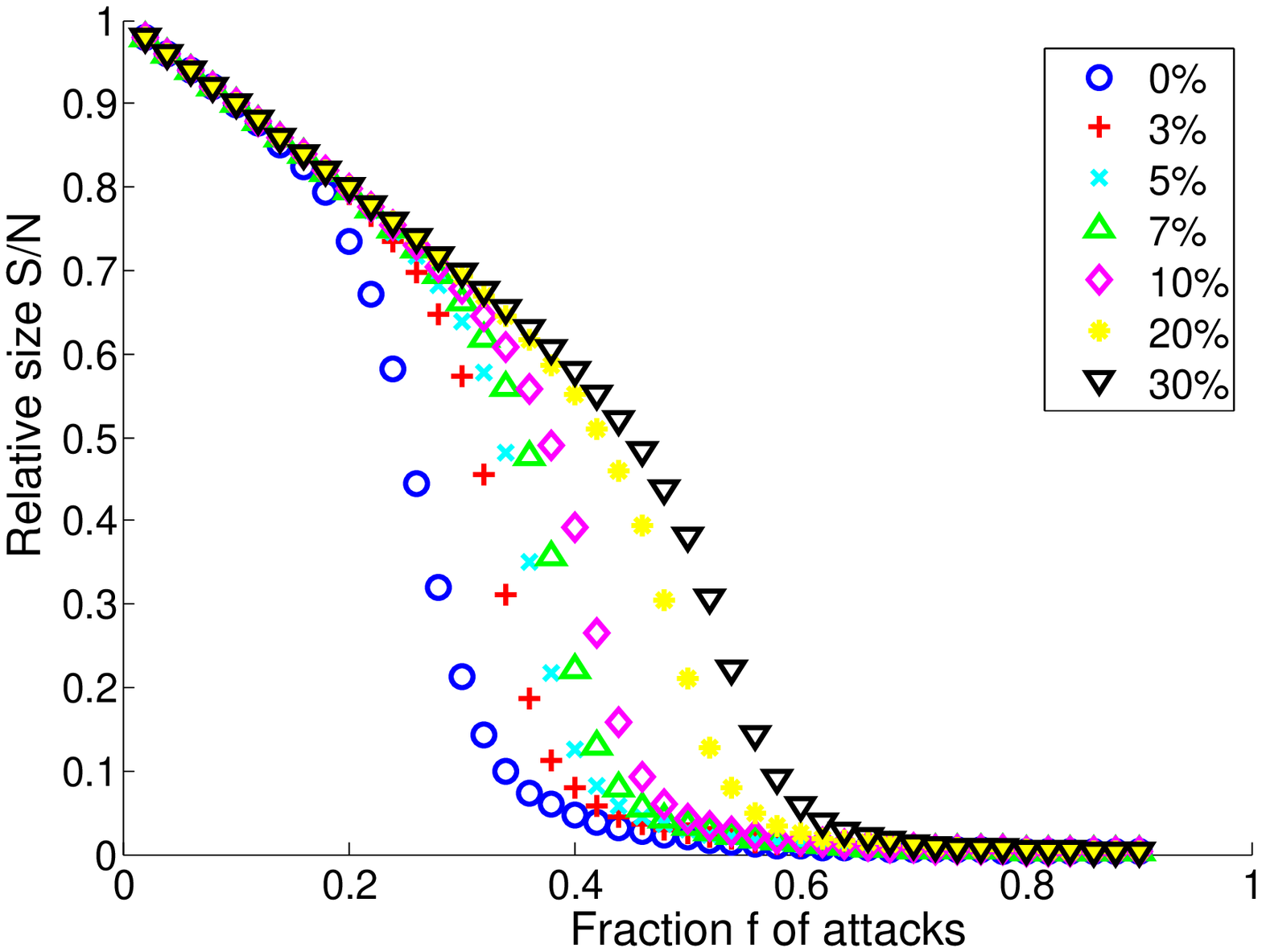}
    \begin{center} (a) DT \end{center}
  \end{minipage}
  \hfill
  \begin{minipage}[htb]{.3\textwidth}
    \includegraphics[height=37mm]{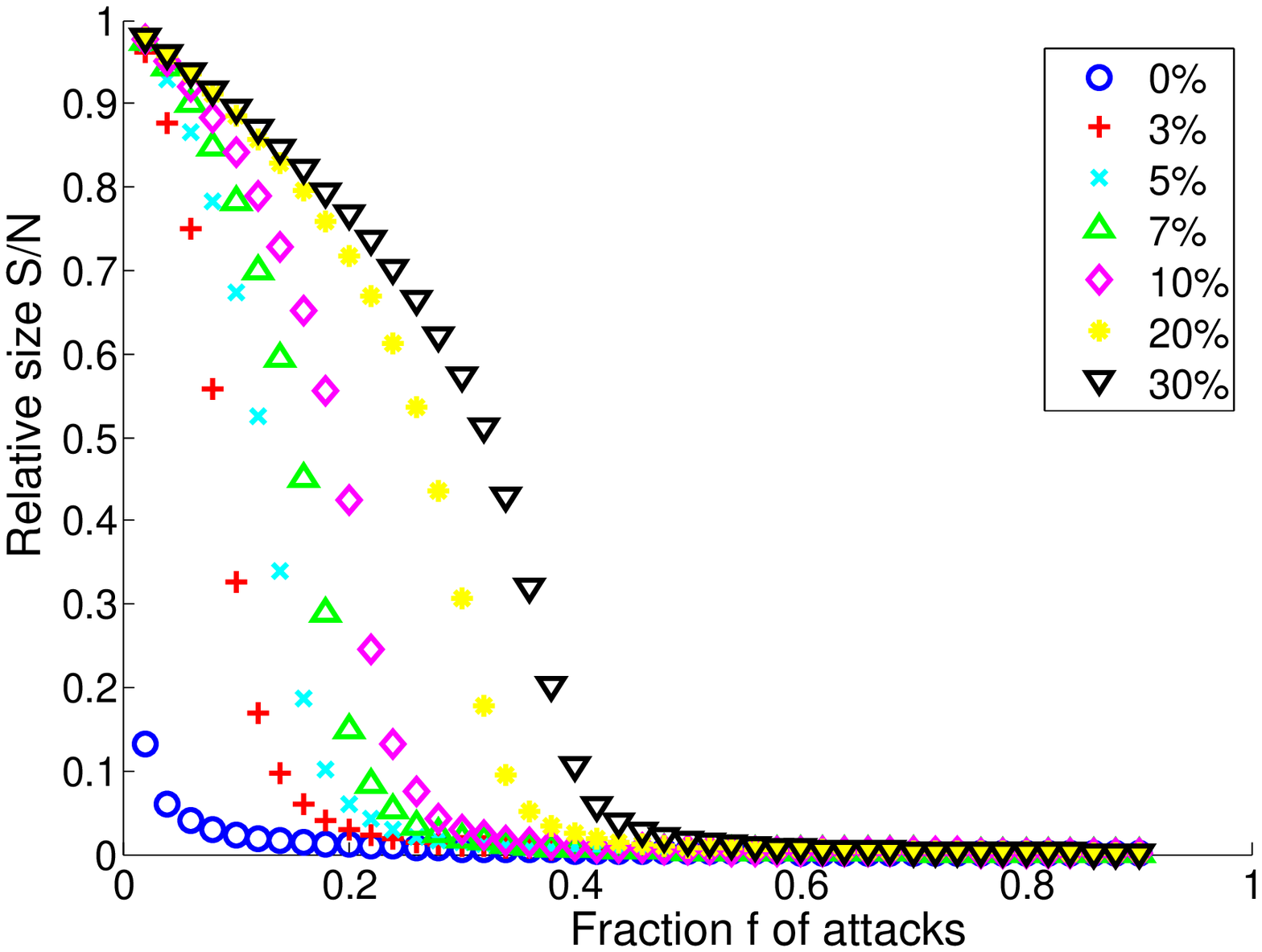}
    \begin{center} (b) RA \end{center}
  \end{minipage}
  \hfill
  \begin{minipage}[htb]{.3\textwidth}
    \includegraphics[height=37mm]{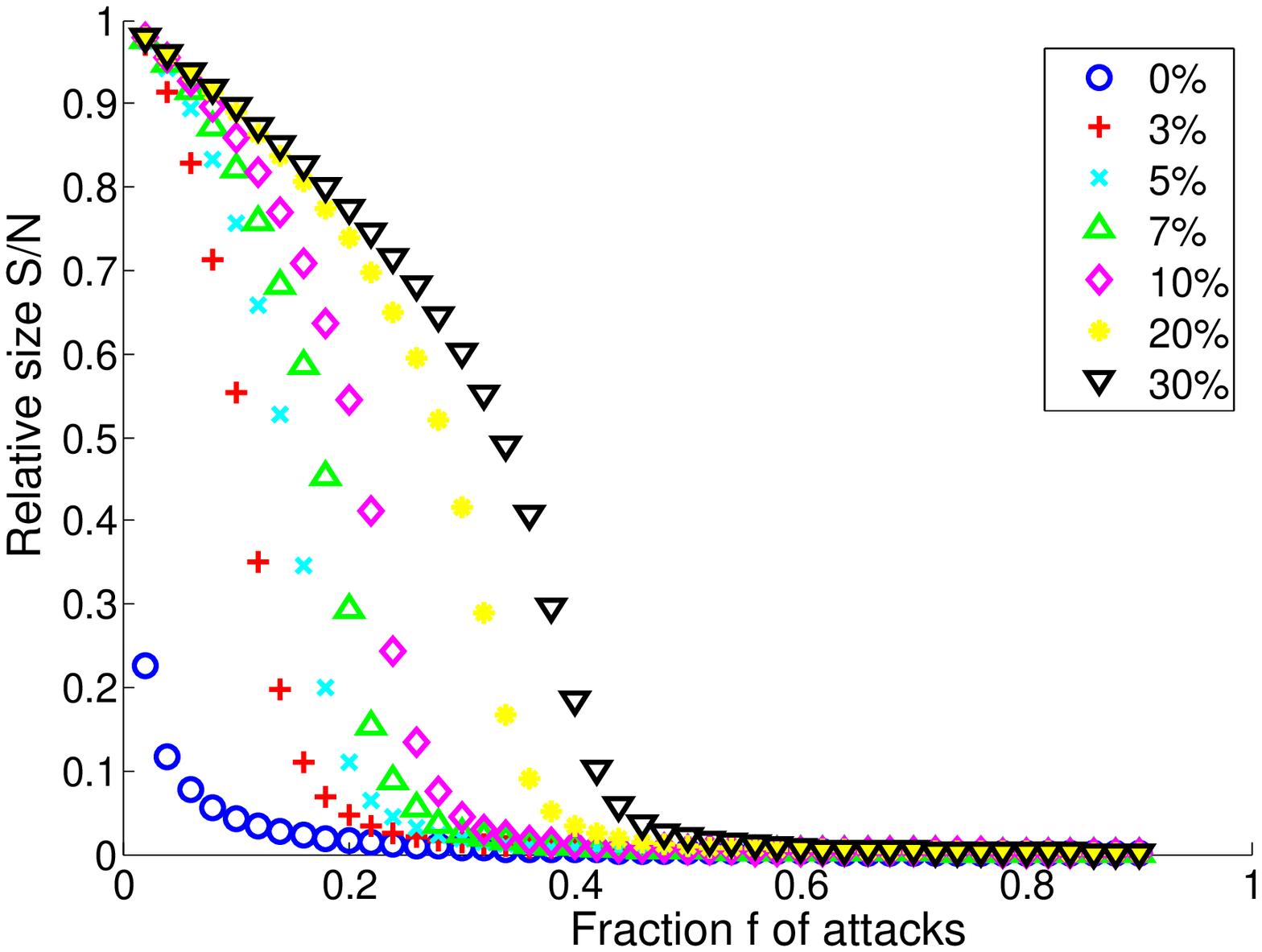}
    \begin{center} (c) DLSF \end{center}
  \end{minipage}
  \caption{(Color online) Relative size $S/N$ of the GC against
 intentional attacks in (a) DT, (b) RA, and (c) DLSF networks
 at the shortcuts rates in legend.}
  \label{fig_GC_attack}
\end{figure}

\begin{figure}[htb]
  \begin{minipage}[htb]{.3\textwidth}
    \includegraphics[height=37mm]{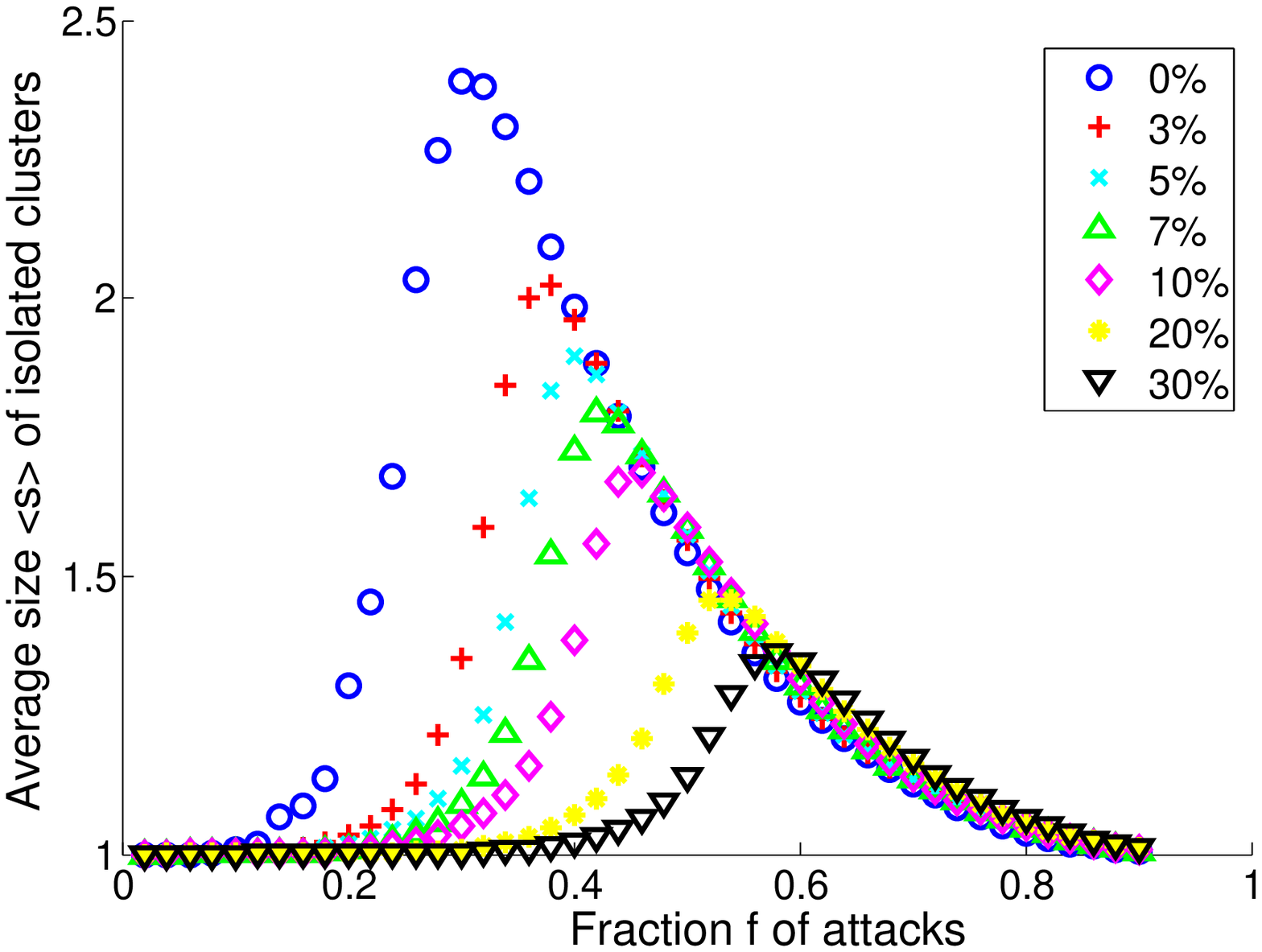}
    \begin{center} (a) DT \end{center}
  \end{minipage}
  \hfill
  \begin{minipage}[htb]{.3\textwidth}
    \includegraphics[height=37mm]{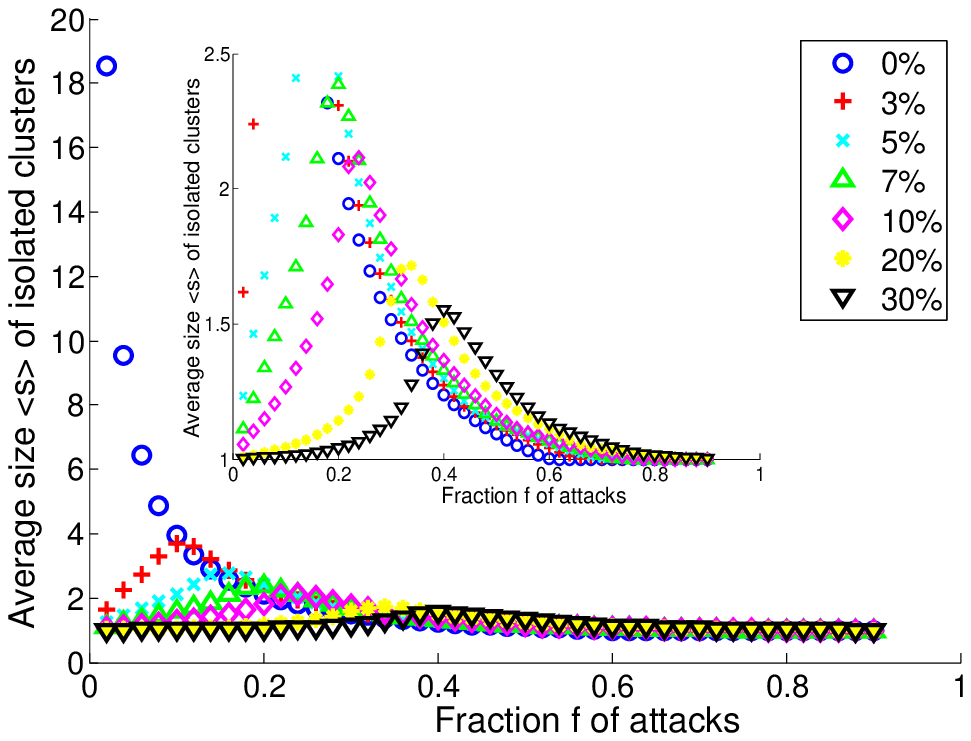}
    \begin{center} (b) RA \end{center}
  \end{minipage}
  \hfill
  \begin{minipage}[htb]{.3\textwidth}
    \includegraphics[height=37mm]{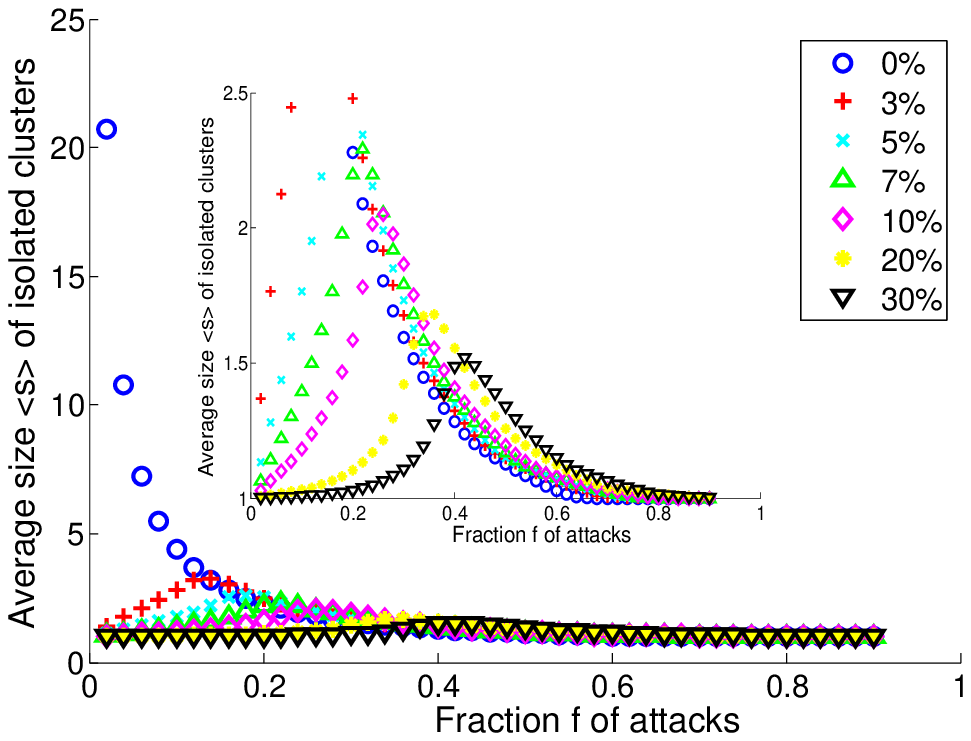}
    \begin{center} (c) DLSF \end{center}
  \end{minipage}
  \caption{(Color online) Average size $\langle s \rangle$
 of isolated clusters except the GC against intentional attacks
 in (a) DT, (b) RA, and (c) DLSF networks
 at the shortcuts rates in legend.
 Inset shows the peaks enlarged by other scale of the vertical axis.}
  \label{fig_size_attack}
\end{figure}

\begin{figure}[htb]
  \begin{minipage}[htb]{.47\textwidth}
    \includegraphics[height=45mm]{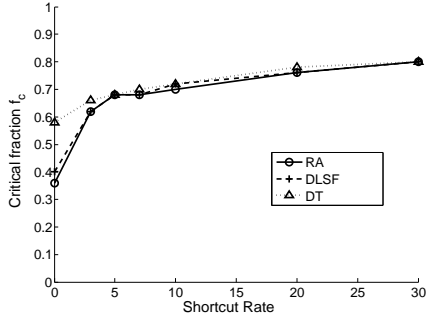}
    \begin{center} (a)  random failures \end{center}
  \end{minipage}
  \hfill
  \begin{minipage}[htb]{.47\textwidth}
    \includegraphics[height=45mm]{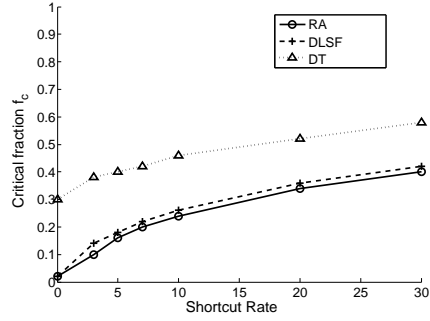}
    \begin{center} (b) intentional attacks \end{center}
  \end{minipage}
 \hfill
  \begin{minipage}[htb]{.47\textwidth}
    \includegraphics[height=45mm]{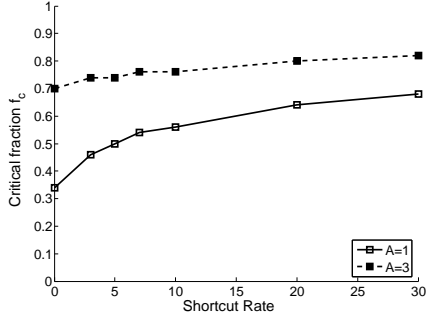}
    \begin{center} (c)  random failures \end{center}
  \end{minipage}
  \hfill
  \begin{minipage}[htb]{.47\textwidth}
    \includegraphics[height=45mm]{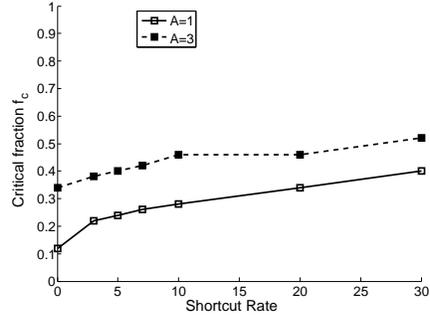}
    \begin{center} (d) intentional attacks \end{center}
  \end{minipage}
  \caption{The critical value $f_{c}$ of removed nodes
 vs. shortcut rate in  (a)(b) RA, DLSF, DT,
 and (c)(d) LESF networks.
  The piece-wise linear lines guide the increasing.}
  \label{fig_fc}
\end{figure}

Figures \ref{fig_GC_attack} and \ref{fig_size_attack}
show the effect of shortcuts on the robustness 
against the targeted attacks on hubs.
In particular, around the shortcuts rate $10 \%$, 
the extremely vulnerable RA and DLSF networks are improved up to the
similar level to DTs.
We compare the critical values of fraction $f_{c}$ of removed nodes 
at the peak of the average size $\langle s \rangle$, 
as the GC is broken off.
As shown in Figure \ref{fig_fc}(b),
the critical values $f_{c}$ in RA and DLSF networks 
reach to $0.3$ at the level of the original DTs without shortcuts.
It is consistent with the effect in evolving networks with local
preferential attachment \cite{Sun06}
that the tolerance becomes higher as increasing the cutoff under the
same average degree $\langle k \rangle$ and size $N$.
We emphasize that, by adding shortcuts around $10 \%$ 
under almost invariant degree distributions, 
the robustness against the intentional attacks 
can be considerably improved up to the similar level to the fully rewired
networks by ignoring the geographical constraints
\cite{Hayashi06b}.

\begin{figure}[htb]
  \begin{minipage}[htb]{.47\textwidth}
    \includegraphics[height=45mm]{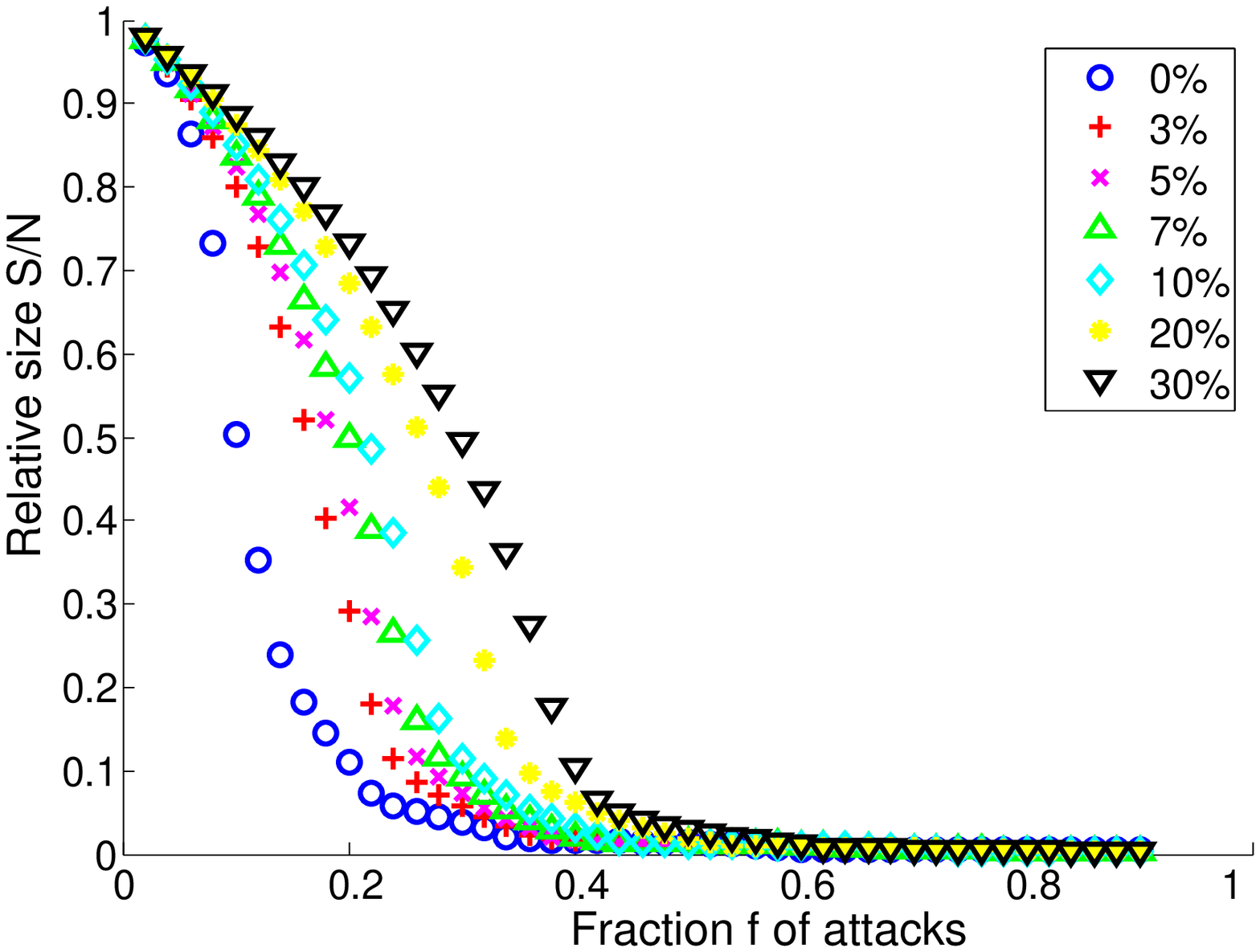}
    \begin{center} (a) $S/N$ at $A=1$ \end{center}
  \end{minipage}
  \hfill
  \begin{minipage}[htb]{.47\textwidth}
    \includegraphics[height=45mm]{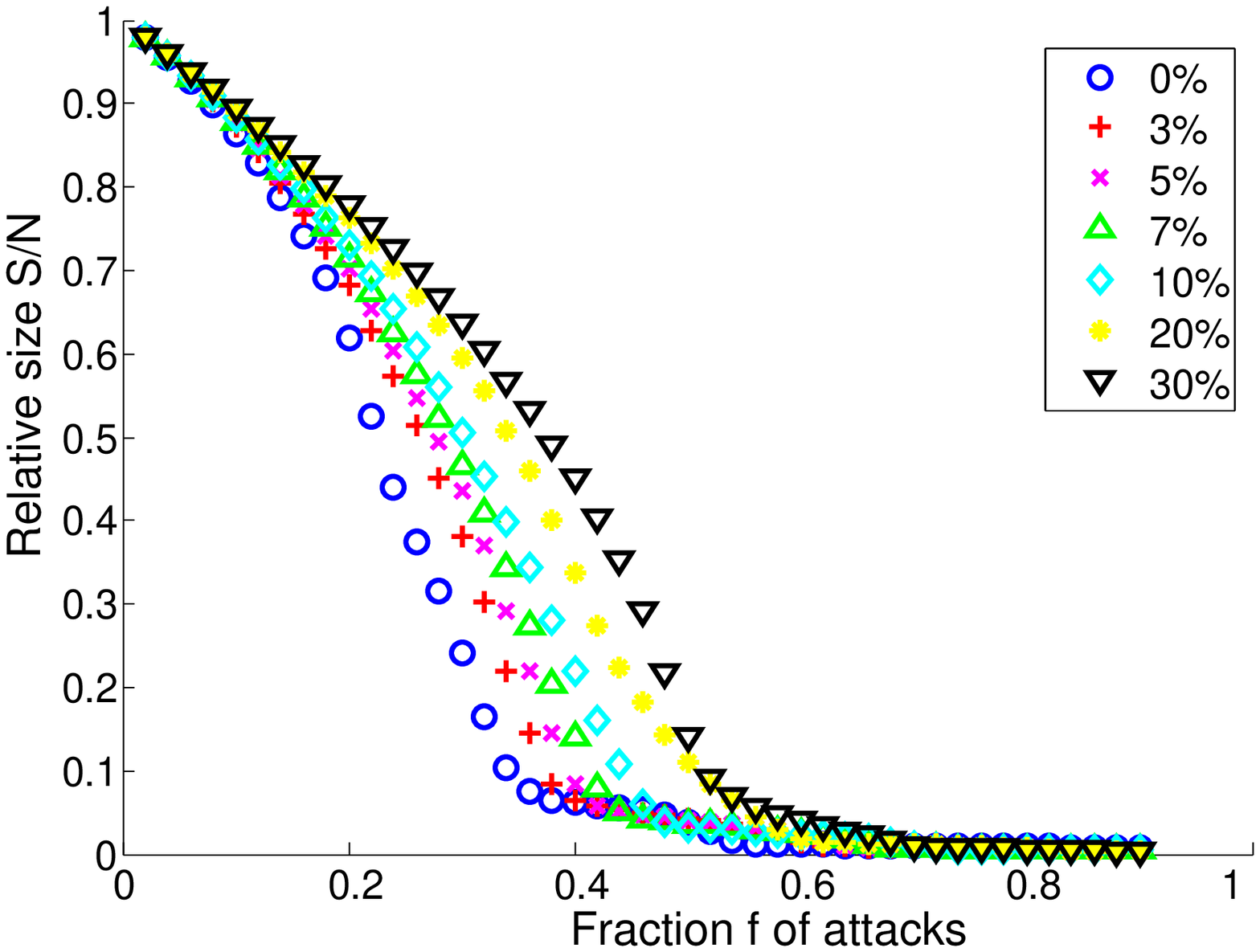}
    \begin{center} (b) $S/N$ at $A=3$ \end{center}
  \end{minipage}
  \hfill
  \begin{minipage}[htb]{.47\textwidth}
    \includegraphics[height=45mm]{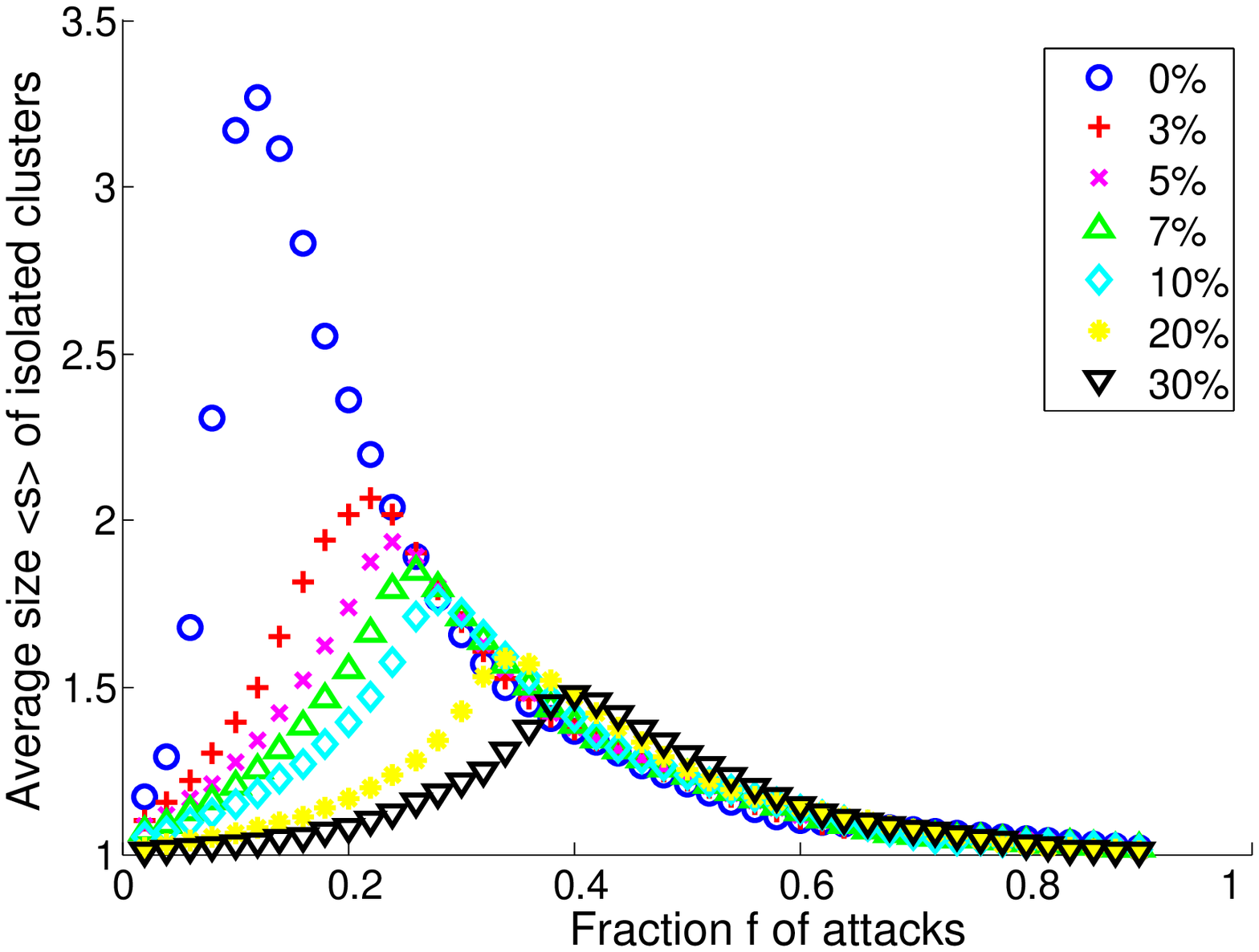}
    \begin{center} (c) $\langle s \rangle$ at $A=1$ \end{center}
  \end{minipage}
  \hfill
  \begin{minipage}[htb]{.47\textwidth}
    \includegraphics[height=45mm]{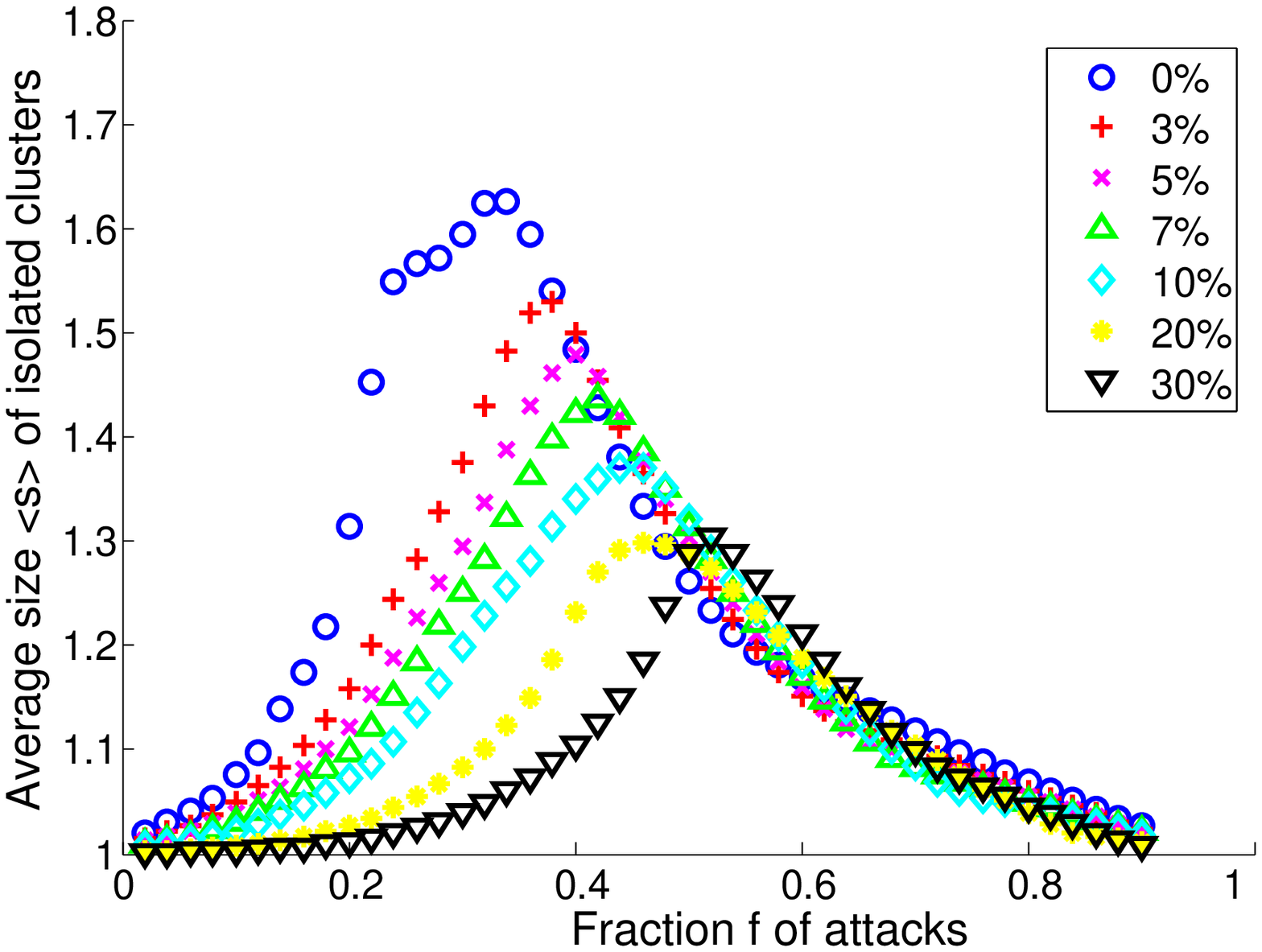}
    \begin{center} (d) $\langle s \rangle$ at $A=3$ \end{center}
  \end{minipage}
  \caption{(Color online) Relative size $S/N$ of the GC against
 intentional attacks in LESF networks at (a) $A=1$ and (b) $A=3$.
 Average size $\langle s \rangle$
 of isolated clusters except the GC against intentional attacks
 in LESF networks at (c) $A=1$ and (d) $A=3$.}
  \label{fig_attack_SFL}
\end{figure}

The effects on LESF networks by adding shortcuts 
are also obtained 
in Fig. \ref{fig_attack_SFL}.
The case of $A = 3$ is more robust because of less geographical
constraint with a larger number $M$ of the total links.
Figures \ref{fig_fc}(c)(d) shows the improvement of the 
critical values $f_{c}$; 
the increase is remarkable in less than the shortcuts rate $10 \%$ 
as similar to Figs. \ref{fig_fc}(a)(b).
These results are also obtained in the LESF networks without the
periodic boundary conditions.

\subsection{Simulation for AS networks}
Historically, in the 1960s, the Internet was motivated to design a 
self-organized computer network with the survival capability for
communication that is highly resilient to local failures
\cite{Satorras04}.
Today, it evolves to one of the world-wide large scale systems, 
whose topology belongs to a SF network with a power-law degree
distribution \cite{Faloutsos99}.
The SF nature of the Internet exhibits both error tolerance and attack
vulnerability \cite{Albert00,Cohen00a,Cohen00b,Satorras04}.
Moreover, the geographical constraints on the topological linkings 
\cite{Yook02,Gastner06} implicitly affect the robustness, 
indeed, the numerical study \cite{Albert00}
has been shown more serious result in the Internet than that in a
relational SF network called Barab\'{a}si-Albert model 
without geographical constraints.
For the realistic case, 
we examine an improvement of the robustness against the attacks in
particular, 
when some shortcuts are virtually added to the Internet.
We use the topology data \cite{CAIDA}
at the level of autonomous system (AS) 
derived from RouteViews BGP table snapshots 
by CAIDA (Cooperative Association for Internet Data Analysis).

\begin{figure}[htb]
  \begin{minipage}[htb]{.47\textwidth}
    \includegraphics[height=50mm]{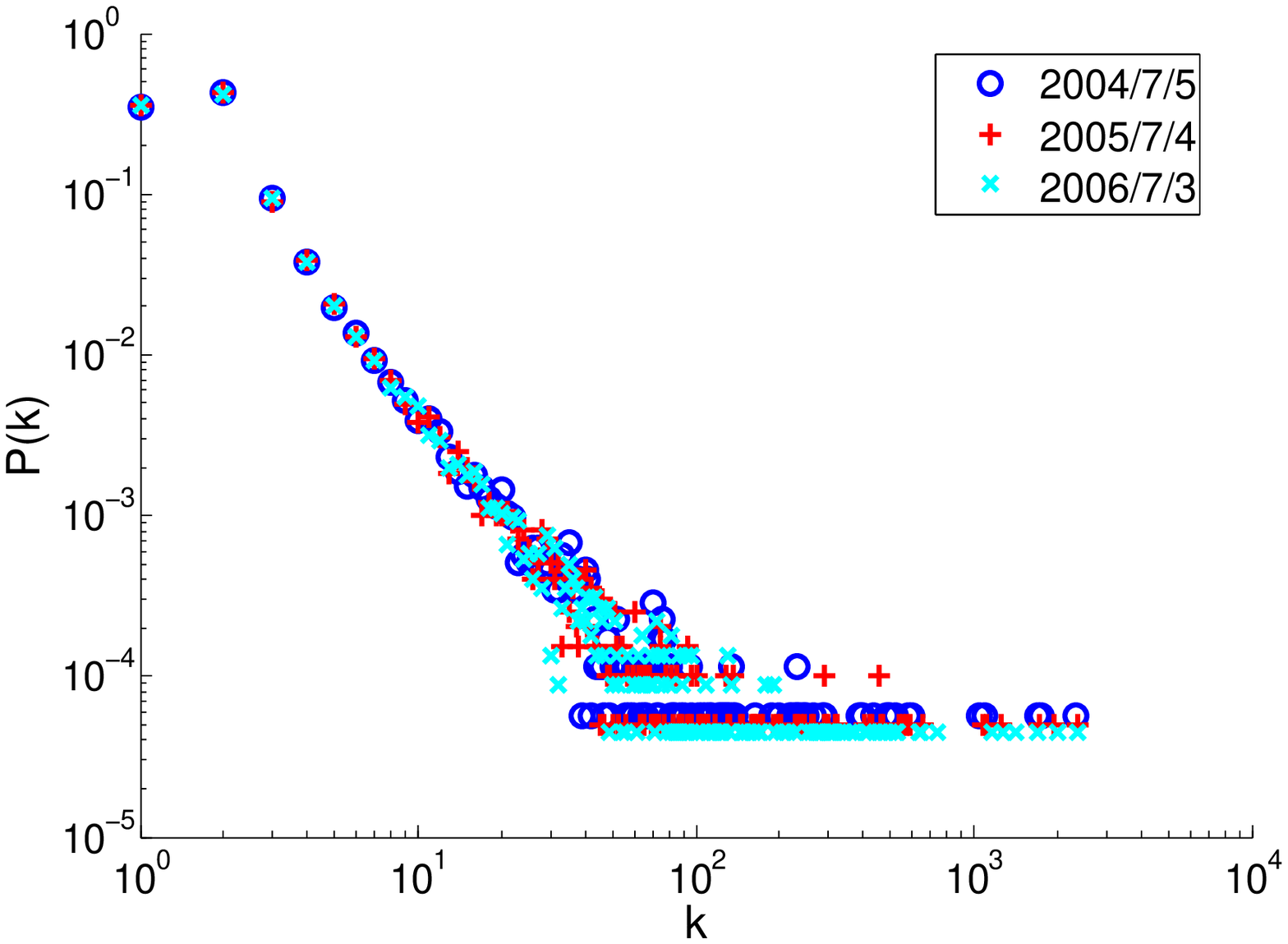}
    \begin{center} (a) $P(k)$ \end{center}
  \end{minipage}
  \hfill
  \begin{minipage}[htb]{.47\textwidth}
    \includegraphics[height=50mm]{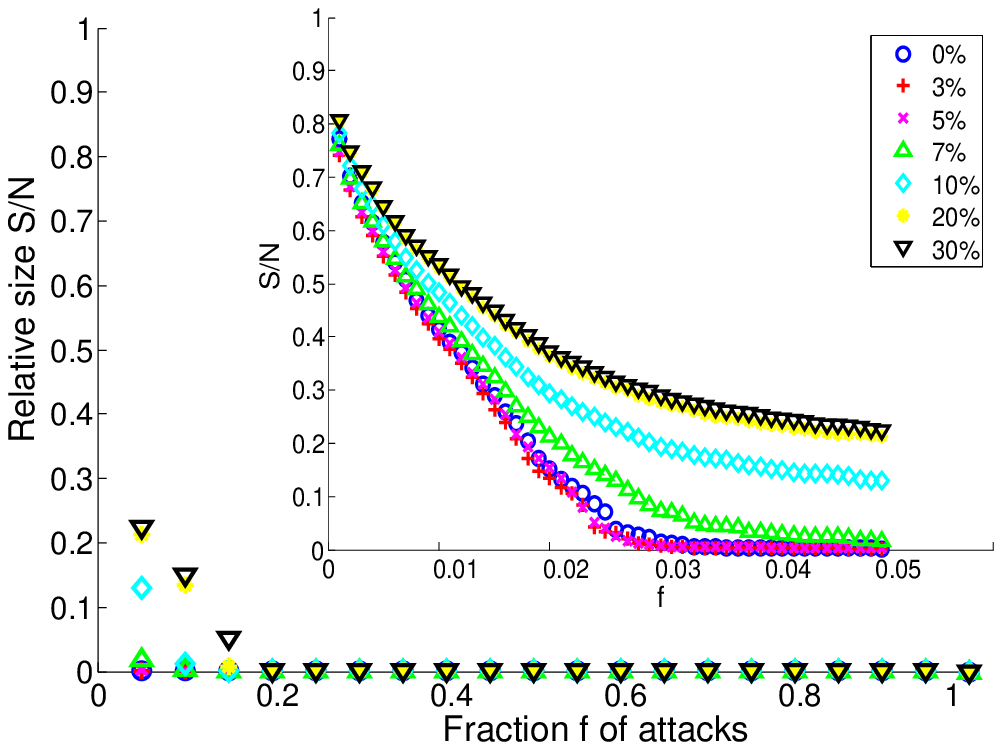}
    \begin{center} (b) GC \end{center}
  \end{minipage}
  \hfill
  \begin{minipage}[htb]{.47\textwidth}
    \includegraphics[height=50mm]{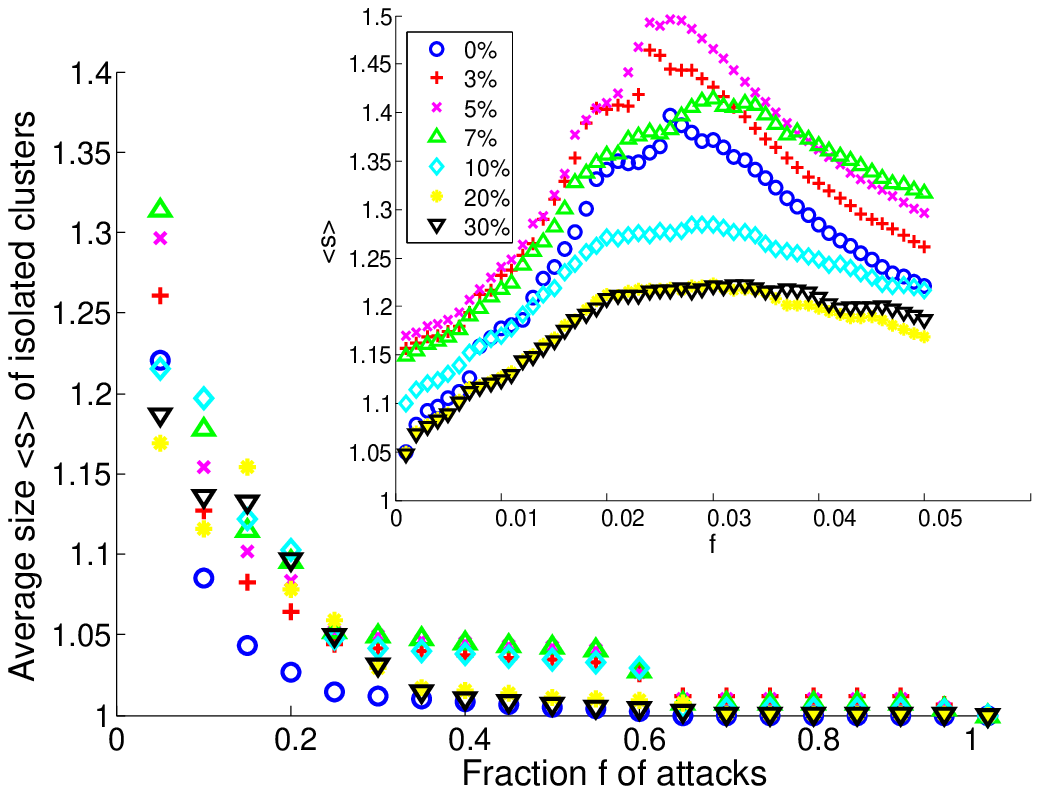}
    \begin{center} (c) $\langle s \rangle$ \end{center}
  \end{minipage}
  \caption{(Color online) Results for the AS networks.
 (a) Degree distribution $P(k)$.
 Each mark corresponds the
 year/month/day.
 (b) Relative size $S/N$ of the GC and
 (c) the average size $\langle s \rangle$ of isolated clusters except
 the GC against intentional attacks on the network in 2006.
 Insets shows the enlarged parts for the small fraction $f$.
 These are similar in other two years.}
  \label{fig_AS_attack}
\end{figure}

Figure \ref{fig_AS_attack}(a) shows a power-law degree distribution in
the AS networks with a few huge hubs.
We also find small deviation $P(k)$ 
for the shortcuts added into these data, 
however the linearity in log-log plot is almost invariant.
Figures \ref{fig_AS_attack}(b)(c) show the effect of shortcuts on the
tolerance of connectivity against the targeted attacks; 
the GC survives even in a double amount of attacks at the breaking of
the original networks without shortcuts, 
and the peak of $\langle s \rangle$ is 
slightly shifted to right.
The breaking around the attack rate $3 \%$ 
is consistent with the previous simulations 
\cite{Albert00,Cohen00a,Satorras04}.
Since a smaller average degree 
$\langle k \rangle$ is improper for maintaining the
connectivity in spite of a small average clustering coefficient
$\langle C \rangle$ 
as shown in Table \ref{table_net_size}, 
these results may be related to a structural vulnerability including
tree-like stubs.

\begin{table}[htb]
\begin{center}
\begin{footnotesize}
\begin{tabular}{c|cc|ccc|c} \hline
Network & $N$ & $M$ & $\langle k \rangle$
& $\langle C \rangle$ & $\langle L \rangle$ & $P(k)$ \\ \hline
DT   & 1000  & 2993 & 5.986 & 0.441 & 9.02 & lognormal \\
RA   & 1000  & 2993 & 5.986 & 0.767 & 4.13 & power-law \\
DLSF & 1000  & 2993 & 5.986 & 0.726 & 4.65 & with cutoff \\ \hline
LESF & & & & & & \\
($A=1$) & 1024 & 1831 & 3.576 & 0.342 & 14.5 & with cutoff \\
($A=3$) & 1024 & 2673 & 5.221 & 0.104 & 4.87 & with cutoff \\ \hline
AS04 & 17509 & 35829 & 4.0926 & 0.234 & 3.77 & power-law \\
AS05 & 19846 & 40485 & 4.0799 & 0.249 & 3.79 & power-law \\
AS06 & 22456 & 45050 & 4.0123 & 0.219 & 3.87 & power-law \\ \hline
\end{tabular}
\end{footnotesize}
\end{center}
\caption{Summary of the topological characteristics: the
 network size $N$, the total number of links $M$,
 the average degree $\langle k \rangle$,
 the average clustering coefficient
 $\langle C \rangle$, the average path length $\langle L \rangle$ 
 based on the minimum number of hops, 
 and the types of degree distribution $P(k)$.} \label{table_net_size}
\end{table}

\section{Conclusion} \label{sec4}
To improve the weakened connectivity 
by cycles in a theoretical prediction \cite{Huang05a,Huang05b},
we investigate effects of shortcuts on the robustness in geographical SF
networks.
Something of randomness 
\cite{Newman00a,Newman00b,Kleinberg00,Hayashi06b}
is expected
to relax the geographical constraints that tend to make cycles locally.
Since many real complex systems belong to SF networks
\cite{Barabasi02,Buchanan02}
and are embedded in a metric space \cite{Yook02,Gastner06},
in addition, 
planar networks are suitable for efficient routings \cite{Bose04}, 
we consider a family of planar SF 
network models called RA \cite{Doye05,Zhou05},
DT \cite{Imai00,Okabe00}, and DLSF \cite{Hayashi06b}, 
a non-planar basic geographical model called LESF \cite{Avraham03}, 
and a real data of the Internet at the AS level \cite{CAIDA}
as an example for the virtual examination.
Our numerical results show that the robustness is improved by shortcuts
around $10 \%$ rate 
maintaining the small distance $\langle D \rangle$ 
and number of hops $\langle L \rangle$ on the optimal paths 
(with respect to the shortest and the minimum number of hops, 
respectively)
in each network, 
under similar degree distributions to the original ones.
In particular, the improvement 
is remarkable in the intentional attacks on hubs.
However, some cases exhibit weak effects
which depend on the values of other 
topological characteristics such as 
 $\langle k \rangle$ and  $\langle C \rangle$.
We will further study for comprehending the properties.
On the other hand, 
these results give an insight for practical constructing of 
a geographical network, since the robustness can be 
effectively increased by adding a small fraction of shortcuts.

\section*{Acknowledgment}
The authors would like to thank CAIDA for using the AS
relationship data \cite{CAIDA}.
This research is supported in part by 
Grant-in-Aide for Scientific Research in Japan, No.18500049.



\begin{figure}[htb]
  \begin{center}
  \includegraphics[width=60mm]{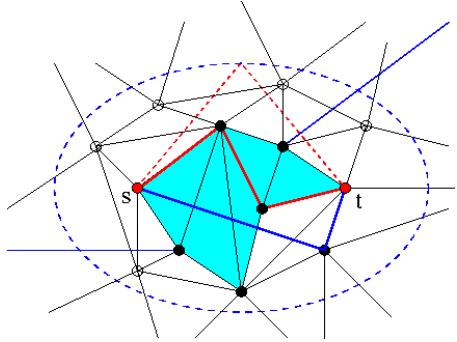}
  \end{center}
  \caption{(Color online) Illustration of the extended routing.
 The (red) thick route on the edges of (cyan) shaded faces
 is the shortest path $l_{s}$ whose distance is the same as the dashed
 (red) chord of the ellipsoid.
 The (blue) thick route is the optimal path including a shortcut.}
  \label{fig_ellipsoid}
\end{figure}

\begin{figure}[htb]
  \begin{center}
  \includegraphics[width=100mm]{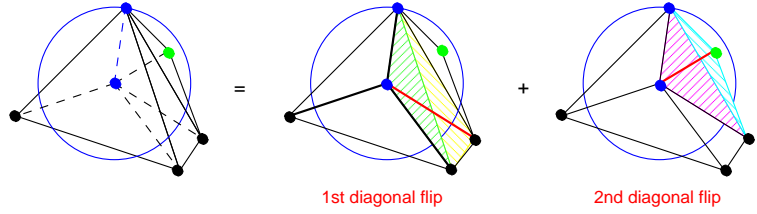}
  \end{center}
  \caption{(Color online) Linking procedures in a Delaunay-like SF network.
 The long-range links (black solid lines in the left)
 are exchanged to red ones in the shaded triangles
 by diagonal flips in the middle and right.
 The dashed lines are new links from the barycenter,
 and form new five triangles with contours in the left
 (The two black solid lines crossed with dashed lines
 are removed after the second diagonal flip).}
  \label{fig_diagonal_flip}
\end{figure}

\section*{Appendix 1}
\appendix
For adding shortcuts, 
the efficient routing algorithm \cite{Bose04} on a planar network can be
extended as follows  (see Figure \ref{fig_ellipsoid})
in the  ellipsoid 
whose chord is defined by the distance of the shortest path
$l_{s}$ on the edges of faces that intersect the straight line between
the source and terminal as the two focuses.
We describe the outline of procedures .
\begin{itemize}
  \item Find the shortest path $l_{s}$ on the original planar network
	without shortcuts.
  \item Then search shorter one including shortcuts in the ellipsoid.
  \item Through backtrackings from the terminal to the source
	in the above process, 
	prune the nodes that located out of the ellipsoid or on longer
	paths than $l_{s}$ by using the positions.
\end{itemize}

We expect the additional steps for searching are not so much as visiting
almost all nodes, 
when the rate of shortcuts is low. Moreover, 
even in this case, the robustness of connectivity 
can be considerably improved.

\section*{Appendix 2}
\appendix
The geographical networks are constructed as follows.

\underline{Planar networks \cite{Hayashi06b}}
\begin{description}
 \item[Step 0:] Set an initial planar triangulation on a space. 
 \item[Step 1:] At each time step, 
	    select a triangle at random and add a new node at the  
            barycenter.
	    For each model, different linking processes are applied.
	    \begin{description}
	     \item[RA:] Then, 
			connect the new node to its three nodes
			as the subdivision of triangle.
	     \item[DLSF:] Moreover, 
			by iteratively applying diagonal flips \cite{Imai00},  
			connect it to the nearest node within a 
			radius defined by the  distance between the new
			node and the nearest node of the chosen
			triangle, as shown in
			Fig. \ref{fig_diagonal_flip}. \\
			If there is no nearest node within the radius, 
			this flipping is skipped, therefore the new node
			is connected to the three nodes.
	     \item[DT:] After the subdivision of the chosen triangle, 
			diagonal flips are globally applied to a
			pair of triangles until the minimum angle is not
			increased by any exchange of diagonal links in a
			quadrilateral. 
	    \end{description}
 \item[Step 2:] The above process is repeated until the required 
            size $N$ is reached. 
\end{description}

\underline{LESF networks \cite{Avraham03}}
\begin{description}
 \item[Step 0:] To each node on the lattice,   
            assign a random degree $k$ taken from the distribution   
            $P(k) = C k^{- \gamma}$, $m \leq k \leq K$, $\gamma > 2$,    
            where   
            $C \approx (\gamma -1) m^{\gamma -1}$ 
	    is the normalization constant for large $K$.       
 \item[Step 1:]  Select a node $i$ at random, and connect it to its 
            nearest neighbors until its connectivity $k_{i}$ is realized, 
            or until all nodes up to a distance 
	    $r(k_{i}) = A k_{i}^{1/d}$
            have been explored:    
            The connectivity quota $k_{j}$ of the target node $j$ is   
            already filled in saturation. Here $A >0$ is a constant.
 \item[Step 2:] The above process is repeated for all nodes.   
\end{description}

\end{document}